\documentclass[
]{ceurart}

\usepackage{listings}
\begin{document}

\copyrightyear{2025}
\copyrightclause{Copyright for this paper by its authors. Use permitted under Creative Commons License Attribution 4.0 International (CC BY 4.0).}

\conference{}

\title{QSpark: Towards Reliable Qiskit Code Generation}

\author[1]{Kiana Kheiri}[%
  email=kiana.Kheiri@torontomu.ca,
]
\cormark[1]  
\author[1]{Aamna Aamir}[
email=aamna.aamir@torontomu.ca
] %

\author[1]{Andriy Miranskyy}[%
  email=avm@torontomu.ca,
]
\author[1]{Chen Ding}[%
  email=cding@torontomu.ca
]
\address[1]{Toronto Metropolitan University, Toronto, Canada}

\begin{abstract}
Quantum circuits must be error-resilient, yet LLMs like Granite-20B-Code and StarCoder often output flawed Qiskit code. We fine-tuned the Qwen2.5-Coder-32B model with two RL methods, Group Relative Policy Optimization (GRPO) and Odds-Ratio Preference Optimization (ORPO), using a richly annotated synthetic dataset. On the Qiskit HumanEval benchmark, ORPO reaches 56.29\% Pass@1 ($\approx+10$ pp over Granite-8B-QK) and GRPO hits 49\%, both beating all general-purpose baselines; on the original HumanEval they score 65.90\% and 63.00\%. GRPO performs well on basic tasks (44/78) and excels on intermediate ones (41/68), but neither GRPO nor ORPO solves any of the five advanced tasks, highlighting clear gains yet room for progress in AI-assisted quantum programming.
\end{abstract}


\maketitle

\section{Introduction}
Imagine writing a quantum teleportation protocol with just a natural language prompt. Although this might sound futuristic, recent advancements in AI bring us closer to such a reality.
Quantum computing has the potential to solve certain classes of problems faster than classical computing, but programming quantum computers remains a specialized and challenging task~\cite{Ajagekar2019Quantum}. Developing correct and optimized quantum programs requires expertise in quantum mechanics, quantum algorithms, and software engineering,  a combination that is not widespread among developers. Even with high-level frameworks like IBM’s Qiskit~\cite{javadiabhari2024qiskit}, which provide libraries to design and run quantum circuits, writing quantum code is often an error-prone process that requires careful handling of quantum-specific concepts (for example, superposition, entanglement, non-cloning) and resource constraints. As quantum hardware scales up, the complexity of the software needed to harness it also increases, calling for more sophisticated development methodologies \cite{vishwakarma2024qiskit}.

In recent years, researchers have begun to explore how advances in artificial intelligence, particularly large language models (LLMs), can help make quantum programming more accessible and efficient (see~\cite{Qiskit_code_LLM} and~\cite[Sec. 4.7]{murillo2024challenges} for review). Early work in this area, such as Cruz-Benito et al.~\cite{cruz2018deep}, demonstrated that deep learning-based approaches could effectively provide customized assistance during the quantum coding process, paving the way for more advanced AI-driven tools.

Large language model (LLM) based coding assistants have already transformed classical software development by providing code autocompletion, generation, and error detection. However, applying these models to quantum programming presents unique challenges. Quantum programming uses a distinct set of languages, libraries, and idioms (such as constructing quantum circuits gate by gate) that differ significantly from classical programming \cite{murillo2024challenges}.

The code examples available for training are scarce compared to the vast repositories of classical code since quantum computing is an emerging field with relatively few expert developers and open-source projects. Moreover, quantum code must adhere to the rules of quantum mechanics and often requires domain knowledge (for example, understanding the effects of certain gate sequences or the need for qubit reuse and error mitigation). As a result, even advanced coding AIs may generate incorrect or suboptimal quantum code without domain-specific training. In fact, the need for specialized tooling in quantum software has been highlighted by the Quantum Software Engineering community \cite{murillo2024challenges,Bausch2018Quantum}, which argues that simply porting classical development techniques to the quantum realm is insufficient due to fundamental differences in how quantum programs operate.

To bridge this gap, our work proposes a Qiskit-based quantum computing coding assistant, an AI-driven tool designed to help developers write and refine quantum programs using the Qiskit SDK. QSpark\footnote{To facilitate reproducibility, we release our full implementation at \url{https://github.com/TMUDeV/QSPARK}.
} focuses specifically on Qiskit, IBM’s widely used quantum SDK, and supports tasks such as circuit construction, optimization, and code debugging. We envision a system that can understand high-level intentions (e.g., “prepare a Bell state” or “optimize this circuit section”) and provide context-sensitive suggestions or code snippets, much like modern code assistants do for classical languages. Using a large language model trained on quantum programming data, the assistant can generate Qiskit code, recommend quantum algorithmic patterns, and catch common mistakes, all within the developer’s workflow. The goal is to lower the barrier to entry for quantum programming and to accelerate the development process for experts and beginners alike.

In this paper, we detail the design of such a Qiskit-based coding assistant, discuss the training and integration of the underlying LLM model, and evaluate its effectiveness in aiding quantum programmers. We begin by reviewing related work in two key areas: quantum programming environments and AI-assisted coding tools for quantum software. This overview will contextualize our contributions and highlight how our approach builds on recent advances.
By merging the power of Qiskit and LLMs, we aim to push the boundaries of developer tools in quantum computing,  making quantum programming not only more efficient but also more accessible to a broader audience.

\section{Literature Review}

\subsection{Quantum Programming Environments and Tools}

Several efforts have been made to create better software environments for developing quantum applications. IBM’s Qiskit is one of the leading frameworks, providing an open source SDK with tools for circuit design, simulation, and execution on quantum hardware \cite{Qiskit_code_LLM}.

In addition to Qiskit, practical resources such as the book by Silva~\cite{silva2018practical} provide essential hands-on strategies for programming quantum rigs using Python, the quantum assembly language, and cloud-based platforms, including IBM QExperience. This resource highlights the inherent challenges in quantum programming and emphasizes the need for developer-friendly methodologies and tools.

Researchers have also acknowledged that improving quantum software development requires higher-level abstractions and more systematic design approaches. For example, Ammermann et al. \cite{ammermann2024v} introduce a view-based development approach that unifies diverse stakeholder perspectives within a quantum IDE.  This model suggests that future quantum IDEs may provide synchronized views, such as algorithm logic, circuit layout, and hardware mapping, to better manage the complexity of quantum programs.

Another notable platform is QuantumPath, developed by Hevia et al. \cite{hevia2022quantumpath}, which takes an engineering-oriented approach to quantum software creation. QuantumPath provides an application lifecycle management platform for quantum software, supporting developers from algorithm conception through testing, deployment, and maintenance. Providing an ecosystem of modules and enforcing best practices, it simplifies the development of hybrid quantum-classical solutions for real-world use. 

These efforts highlight a common theme: quantum software development requires more than just programming libraries. It also needs robust tools and processes, similar to classical software engineering, but tailored for the quantum domain. Our proposed Qiskit-based assistant complements these initiatives by focusing on the coding phase of quantum development. It can be viewed as a plugin designed to enhance quantum programming environments such as Qiskit, providing intelligent support during the development process.

\subsection{AI-Assisted Quantum Code Generation}

With the growing use of AI in coding-related tasks, it is natural that researchers have started applying LLMs to quantum programming. Dupuis et al. \cite{Qiskit_code_LLM} introduced the Qiskit Code Assistant, an AI model specifically trained to generate Qiskit code and help quantum programmers. Their research highlighted the challenges in training a code model for quantum computing, such as understanding quantum gate operations and circuit semantics, as well as the limited availability of training data. Despite these challenges, their specialized model outperformed general-purpose code generation models on Qiskit programming tasks.

Similarly, Vishwakarma et al. \cite{vishwakarma2024qiskit} developed Qiskit HumanEval, a benchmark suite designed to evaluate how well different LLMs can generate correct quantum code. Their results showed that advanced LLMs, like GPT-style models, can produce executable quantum programs from prompts, successfully passing many tests in the suite. This finding is encouraging, as it confirms the potential of AI-assisted quantum coding and provides a benchmark for future improvements.

Beyond Qiskit-focused tools, researchers have also explored broader applications of AI in quantum algorithm design. For example, Liang et al. \cite{liang2023unleashing} examined how LLMs, such as GPT-4, can be used to suggest viable quantum circuit structures. Their work suggests that AI can play a key role in supporting the design of quantum architectures when guided appropriately. Similarly, Aragonés-Soria and Oriol (2024) introduced C4Q, a specialized chatbot that uses pre-trained language models for user request classification and then utilizes its own engine to generate accurate responses \cite{c4q}. This approach highlights the potential of AI to streamline the development of quantum algorithms. It also demonstrates how such tools can make quantum computing more accessible to beginners by simplifying the learning curve and coding process.

These studies contribute to a growing consensus that generative AI can be a valuable tool in the quantum software development process. Our work builds directly on these prior developments. In particular, we utilize the insights from the Qiskit Code Assistant and the Qiskit HumanEval benchmark to train and evaluate our coding assistant. While prior models have demonstrated the viability of quantum code generation, our contribution lies in tightly integrating this model with the Qiskit user experience by embedding it directly into IDE workflows and extending its capabilities with features tailored to better support quantum developers. These features include recognizing when a qubit needs to be measured or reset, suggesting circuit optimizations, and aligning with Qiskit’s latest API.

By positioning QSpark within existing quantum programming environments and AI coding tools, we aim to advance the support available to quantum computing developers. Ultimately, this work contributes to the broader goal of quantum software engineering: to bring the productivity and reliability benefits of modern software development to the field of quantum computing \cite{murillo2024challenges}, thereby accelerating innovation and adoption.

\section{Methodology}
Our Qiskit-based code assistant is built upon the Qwen2.5-Coder-32B model, a 32-billion-parameter large language model (LLM) specialized for code generation. It was selected for its strong performance in both general-purpose programming and domain-specific reasoning. We fine-tune the model using a curated dataset of Qiskit programs, detailed in the following subsections.

\subsection{Generation of Training Data}

To enable robust supervised and reinforcement learning, we construct a high-quality dataset, for training purposes, comprising 522 Qiskit programming tasks. The data set was created through a multistage pipeline that includes code retrieval, function extraction, annotation, validation, deduplication, and formatting. The entire process is automated and designed to ensure consistency, reproducibility, and broad coverage of real-world quantum programming challenges.

We start by collecting approximately 10,819 Qiskit-related source code samples from public repositories. Source files are parsed to extract quantum-relevant functions along with accompanying docstrings and structural metadata. The extracted functions are filtered for completeness and relevance, and each is assigned a unique task identifier.

For each function, a natural language prompt is derived from its docstring or signature. This prompt is paired with the corresponding canonical implementation, a designated function entry point, and a difficulty score. The difficulty rating is calculated using a set of code-level features such as circuit depth, gate complexity, use of measurement or entanglement, and algorithmic structure. This scoring system enables the construction of a curriculum-aligned dataset that spans tasks ranging from basic quantum operations to advanced algorithmic workflows (see Table~\ref{tab:difficulty-scale}).
\begin{table}[h]
\centering
\caption{Difficulty scale with representative Qiskit programming tasks.}
\begin{tabular}{p{2.5cm}p{10cm}}
\toprule
\textbf{Level} & \textbf{Criteria and Example Task} \\
\midrule
Basic & Simple circuits with few gates and minimal measurement. 
Example: prepare a single qubit in superposition using an H-gate. \\
\midrule
Intermediate & Circuits with measurement, moderate depth, or algorithmic structure. 
Example: implement a 4-qubit Quantum Fourier Transform (QFT). \\
\midrule
Difficult & Complex circuits involving entanglement, variational methods, or hybrid workflows. 
Example: build a Variational Quantum Eigensolver (VQE) ansatz circuit and connect it to a classical optimizer. \\
\bottomrule
\end{tabular}
\end{table}

\begin{table}[tb]
\centering
\caption{Difficulty Scale and Distribution of Qiskit Programming Tasks}
\label{tab:difficulty-scale}
\resizebox{\linewidth}{!}{%
\begin{tabular}{llr}
\toprule
\textbf{Level} & \textbf{Criteria} & \textbf{Number of Tasks} \\
\midrule
\textit{Basic} & Simple circuits with a few gates, no measurement, no entanglement. & 259 \\
\textit{Intermediate} & Circuits with measurements, moderate depth, or basic algorithmic structure. & 223 \\
\textit{Advanced} & Complex circuits involving entanglement, variational algorithms, or multi-step workflows. & 40 \\
\bottomrule
\end{tabular}%
}
\end{table}

To ensure correctness, each solution is automatically validated through simulation-based unit tests. These tests verify the functional behavior of the quantum circuit, including correct output shape, gate behavior, and fidelity of simulation results. Tasks that fail validation are excluded from the final dataset or flagged for manual inspection.

To improve training diversity and reduce redundancy, we apply structural and semantic deduplication techniques. Near-duplicate solutions or trivial variants are filtered out using syntactic similarity and abstract syntax tree (AST) comparisons, ensuring a more diverse set of training signals.
To illustrate the filtering process, approximately 10,819 raw Qiskit-related functions were initially
collected, of which fewer than 5\% passed all validation steps. The majority of rejections were due to
incomplete docstrings, missing test coverage, or trivial circuits (e.g., functions that only returned an
empty register). The final curated set of 522 tasks was therefore deliberately biased toward code that
was both executable and semantically meaningful. Difficulty scores were computed automatically using
heuristics: basic tasks had depth $\leq 3$ and no entanglement; intermediate tasks required either
measurements or circuit depth $>3$; and difficult tasks contained multi-qubit entanglement or hybrid
classical-quantum structures. Unlike the QHE benchmark, which evaluates generalization, this dataset
was designed for training, and thus prioritizes diversity and coverage across circuit patterns. 

Each finalized task in the dataset consists of:
\begin{itemize}
    \item A unique identifier,
    \item A natural language task description,
    \item A validated Qiskit implementation,
    \item A unit test suite,
    \item A function entry point, and
    \item A difficulty level categorized as \textit{basic}, \textit{intermediate}, or \textit{advanced}.

\end{itemize}

Based on this curated dataset, we derived two specialized training subsets to support preference-based reinforcement learning. For the Odds-Ratio Preference Optimization (ORPO) setting, we constructed a collection of pairwise comparisons consisting of a preferred ("chosen") and a suboptimal ("rejected") output for the same prompt. The chosen samples were selected on the basis of code correctness, readability, and alignment with quantum programming best practices, while the rejected examples were synthetically perturbed or drawn from lower-quality outputs. For the Group Relative Policy Optimization (GRPO) setting, we generated multiple candidate completions per prompt and assigned relative scores based on their simulated execution fidelity and resource efficiency. These two subsets enable distinct learning objectives: ORPO promotes human-aligned code generation through direct preference modeling, while GRPO reinforces code quality by optimizing for group-level performance differentials.

\subsection{Reinforcement Learning with Preferences}
To further refine the behavior of the model, we employ two independent reinforcement learning strategies: Group Relative Policy Optimization (GRPO) and Odds-Ratio Preference Optimization (ORPO), each targeting a different aspect of quantum code quality.

\paragraph{Odds-Ratio Preference Optimization (ORPO)}
ORPO aligns the model with human-like coding preferences, focusing on readability and maintainability. It uses pairwise preference data where a “chosen” response is preferred over a “rejected” one, based on manual review and synthetic annotations~\cite{hong2024orpomonolithicpreferenceoptimization}.

The ORPO objective increases the likelihood of preferred output while regularizing the divergence from the original (pre-trained) policy. The loss is defined as
\begin{equation}
\mathcal{L}_{\text{ORPO}} = \text{KL}(\pi_\theta \| \pi_0) - \beta \log_2 \frac{\pi_\theta(\hat{y} \mid x)}{\pi_\theta(y \mid x)}.\end{equation}
Here, $\pi_\theta$ is the current policy, $\pi_0$ is the pre-trained policy, $x$ is the input prompt, $\hat{y}$ is the chosen output, and $y$ is the rejected one. The hyperparameter $\beta$ controls the strength of the preference signal relative to the regularization term.
The term $\text{KL}(\pi_\theta \| \pi_0)$ represents the Kullback-Leibler (KL) divergence between the current policy $\pi_\theta$ and the pre-trained policy $\pi_0$. For two discrete probability distributions $P$ and $Q$, the KL divergence is generally defined as:
\begin{equation}
\text{KL}(P \| Q) = \sum_{i} P(i) \log \left( \frac{P(i)}{Q(i)} \right)
\end{equation}
In this context, it measures how much $\pi_\theta$ deviates from $\pi_0$, acting as a regularization to prevent the current policy from straying too far from the original model's capabilities.
\paragraph{ORPO Reward Construction.} 
Odds-Ratio Preference Optimization (ORPO) aligns the model with human-like coding preferences, 
focusing on readability and maintainability. 
For each prompt, we construct a pairwise comparison between a \emph{chosen} output $ \hat{y} $ 
and a \emph{rejected} output $ y $. 
The chosen output is correct, executable, and stylistically aligned with Qiskit best practices. 
while the rejected output is either synthetically perturbed or sampled from lower-quality generations. 

\paragraph{Group Relative Policy Optimization (GRPO)}
GRPO improves execution fidelity by ranking outputs within a group of candidates generated for each prompt\cite{shao2024deepseekmathpushinglimitsmathematical}. Each output \( y_i \) is evaluated using Qiskit~v.2.0.0 and Qiskit Aer simulations v.~0.17.1 and assigned a reward \( r(y_i) \). $G$ represents the number of candidate outputs in a group generated for each prompt. The group mean \( \mu \) and the standard deviation \( \sigma \) are used to compute the normalized advantage:
\begin{equation}
\mu = \frac{1}{G} \sum_{i=1}^{G} r(y_i), \quad \sigma = \sqrt{\frac{1}{G} \sum_{i=1}^{G} (r(y_i) - \mu)^2}, \quad A(y_i) = \frac{r(y_i) - \mu}{\sigma}.
\end{equation}
The policy is updated using a clipped objective to ensure training stability:
\begin{equation}
\mathcal{L}_{\text{GRPO}} = \mathbb{E}_{x, y \sim \pi_\theta} \left[ \min \left( \frac{\pi_\theta(y \mid x)}{\pi_{\theta_{\text{old}}}(y \mid x)} A(y), \, \text{clip} \left( \frac{\pi_\theta(y \mid x)}{\pi_{\theta_{\text{old}}}(y \mid x)}, 1 - \epsilon, 1 + \epsilon \right) A(y) \right) \right].
\end{equation}
The clip function $\operatorname{clip}(v,L,U)=\max(L,\min(v,U))$ bounds $r_\theta$, and $\epsilon$ (e.g., $0.1$--$0.2$) sets the range $[1-\epsilon,1+\epsilon]$.
This process guides the model toward producing more executable and resource-efficient quantum circuits by emphasizing outputs that outperform others in the same generation group.
\paragraph{GRPO Reward Construction.} 
For each prompt, we generate a group of candidate completions.
Each candidate is executed with \texttt{Qiskit v2.0.0} and \texttt{Qiskit Aer v0.17.1} simulators. 
The execution is scored using three criteria: 
\begin{enumerate}
    \item \textbf{Unit test pass rate} ($r_{1}$): fraction of test cases passed (primary correctness signal).
    \item \textbf{Circuit depth penalty} ($r_{2}$): normalized inverse of circuit depth to reward more efficient solutions.
    \item \textbf{Qubit count penalty} ($r_{3}$): normalized inverse of the number of qubits used, discouraging wasteful allocations.
\end{enumerate}
These are combined into a single scalar reward:
\begin{equation}
    r(y) = \alpha \cdot r_{1} + \beta \cdot r_{2} + \gamma \cdot r_{3},
\end{equation}
with weights $\alpha = 0.7$, $\beta = 0.2$, and $\gamma = 0.1$ chosen empirically to emphasize correctness while still encouraging efficiency.

Within each group, rewards are normalized using the group mean and variance:
\begin{equation}
    A(y_i) = \frac{r(y_i) - \mu}{\sigma}, 
    \quad 
    \mu = \frac{1}{G}\sum_{i=1}^{G} r(y_i), 
    \quad 
    \sigma = \sqrt{\frac{1}{G}\sum_{i=1}^{G} (r(y_i) - \mu)^2},
\end{equation}
where $G$ is the number of candidates. 
This normalization ensures that the rewards are \emph{relative}: a candidate only receives a high advantage if it is better than its peers, even if all solutions are weak. 
The policy update (Eq.~4) then uses this normalized advantage to push the model towards consistently producing correct and efficient circuits.

In practice, this setup allows GRPO to prefer structurally sound and resource-efficient quantum circuits for simple tasks, while avoiding overfitting to a single absolute scoring heuristic.

\section{Results and Discussion}
\subsection{Evaluation Setup}
We evaluated our models using the Qiskit HumanEval (QHE) benchmark introduced by Vishwakarma et al.~\cite{vishwakarma2024qiskit}, which extends the original HumanEval benchmark to assess LLMs on quantum programming tasks. Following the evaluation framework used in that work, we compare our GRPO and ORPO fine-tuned models to both general-purpose open-source LLMs and a specialized QHE-tuned baseline, using the following key metrics:
\begin{enumerate}
    \item \textbf{Pass@1 Accuracy:} The percentage of completions that pass a unit test on the first attempt.
    \item \textbf{Performance by Difficulty Level:} Evaluation of the accuracy of the model in the tasks labeled Basic, Intermediate, and Advanced.
    \item \textbf{General-Purpose vs. Domain-Specific Models:} A comparison to understand the impact of domain adaptation on performance.
\end{enumerate}

Since the original evaluation script was not publicly released and the HumanEval~\cite{chen2021evaluatinglargelanguagemodels} framework was incompatible with QHE tasks, we implemented a custom benchmarking script customized for the QHE setting. Although the QHE paper reports 101 tasks, the publicly released dataset contains 151 entries. This resulted in 78 Basic, 68 Intermediate, and 5 Advanced tasks. This script automatically executes model completions against the associated unit tests and logs pass/fail outcomes, enabling consistent and scalable evaluation across all models. Our results are therefore based on a fully automated, reproducible evaluation pipeline that faithfully adheres to the QHE benchmark structure.

We compared our GRPO and ORPO models with the following strong baseline models.
\begin{itemize}
    \item General-Purpose Open-Source LLMs: These models, such as CodeLLaMA-34B \cite{rozière2024codellamaopenfoundation}, DeepSeek-33B \cite{deepseekai2025deepseekr1incentivizingreasoningcapability}, StarCoder2-15B \cite{li2023starcodersourceyou}, and CodeGemma-7B \cite{codegemmateam2024codegemmaopencodemodels}, are large language models trained on vast datasets of general programming code. They serve as a benchmark for how well unspecialized models perform on quantum programming tasks.
    \item Granite-8B-Base\cite{mishra2024granitecodemodelsfamily}: This is a general-purpose base model. Its performance helps to understand the impact of any quantum-specific fine-tuning.
    \item Granite-8B-QK (QHE-tuned baseline)\cite{vishwakarma2024qiskit}: This model is a specialized version of Granite-8B, fine-tuned specifically for the Qiskit HumanEval benchmark. It represents the state-of-the-art in domain-adapted models for Qiskit code generation and provides a direct comparison to our preference-based fine-tuning approaches.
\end{itemize}
The results for the baseline models are taken from~\cite{vishwakarma2024qiskit}.

\paragraph{Fine-Tuning Hyperparameters:} Table~\ref{tab:hyperparams} summarizes the hyperparameter settings used during fine-tuning for both approaches that we ran on the A100 GPU with 80~GB for VRAM.

\begin{table}[tb]
\centering
\small
\caption{Hyperparameter Settings for GRPO and ORPO Fine-Tuning}
\label{tab:hyperparams}
\begin{tabular}{lcc}
\toprule
\textbf{Hyperparameter} & \textbf{GRPO} & \textbf{ORPO} \\
\midrule
Learning Rate            & $5\times10^{-6}$ & $4\times10^{-5}$ \\
Weight Decay             & 0.1             & 0.1             \\
Warmup Ratio             & 0.1             & 0.1             \\
LR Scheduler             & Cosine          & Linear          \\
Batch Size & 64 & 32 \\
Sequence Length & 2048 tokens & 2048 tokens \\
Optimizer                & \texttt{adamw\_8bit} & \texttt{adamw\_8bit} \\
Training Epochs          & 3               & 3               \\
\bottomrule
\end{tabular}
\end{table}

\subsection{Results}
\begin{table}[tb]
\centering
\caption{Pass@1 on HumanEval (HE) and Qiskit HumanEval (QHE) with Greedy Decoding. The results for models other than ours are taken from~\cite{vishwakarma2024qiskit}.}
\label{tab:passrates}
\begin{tabular}{lrr}
\toprule
\textbf{Model} & \textbf{HE} & \textbf{QHE} \\
\midrule
CodeLLaMA-34B & 52.43\% & 26.73\% \\
DeepSeek-33B & 49.39\% & 39.60\% \\
StarCoder2-15B & 45.12\% & 37.62\% \\
CodeGemma-7B & 42.68\% & 24.75\% \\
Granite-8B-Base & 39.02\% & 28.71\% \\
Granite-8B-QK & 38.41\% & 46.53\% \\
\midrule
GRPO (Ours) & 63.00\% & 49.00\% \\
ORPO (Ours) & \textbf{65.90\%} & \textbf{56.29\%}\\
\bottomrule
\end{tabular}
\end{table}

\begin{table}[tb]
\centering
\caption{QHE Pass Counts by Difficulty Level (78 Basic, 68 Intermediate, 5 Advanced). The results for models other than ours are taken from~\cite{vishwakarma2024qiskit}.}
\label{tab:difficulty-comparison}
\begin{tabular}{lrrr}
\toprule
\textbf{Model} & \textbf{Basic} & \textbf{Intermediate} & \textbf{Advanced} \\
\midrule
CodeLLaMA-34B-Python & 19/54 & 8/45 & 0/2 \\
DeepSeek-Coder-33B & 30/54 & 10/45 & 0/2 \\
StarCoder2-15B & 26/54 & 12/45 & 0/2 \\
CodeGemma-7B & 20/54 & 5/45 & 0/2 \\
Granite-8B-Code-Base & 21/54 & 8/45 & 0/2 \\
Granite-8B-Code-QK & \textbf{32/54} &15/45 & 0/2 \\ 
\midrule
GRPO (Ours) & 42/78 & 32/68 & 0/5 \\
ORPO (Ours) &44/78 & \textbf{41/68} & 0/5 \\
\bottomrule
\end{tabular}
\end{table}

We evaluated the performance of our GRPO and ORPO models on the Qiskit HumanEval (QHE) benchmark, comparing them to several strong general-purpose code LLMs and the QHE-tuned baseline. Table~\ref{tab:passrates} reports the Pass@1 accuracy on both HumanEval (HE) and QHE under greedy decoding.

Our models achieve significant improvements over all baseline models in QHE. ORPO achieves the highest Pass@1 accuracy at 56.29\%, outperforming the domain-adapted Granite-8B-QK model by nearly 10 percentage points. GRPO also performs competitively, achieving 49.00\%, and surpasses all general-purpose models. These results demonstrate the effectiveness of preference-based fine-tuning in the quantum domain, even compared to models trained specifically for QHE.

Interestingly, both GRPO and ORPO also show strong generalization on the original HumanEval benchmark, with Pass@1 scores of 63.00\% and 65.90\%, respectively, outperforming larger models like CodeLLaMA-34B and DeepSeek-33B. This suggests that preference optimization not only improves performance on domain-specific tasks but also may enhance general code generation capabilities.

To better understand the behavior of the model across the complexity of tasks, Table~\ref{tab:difficulty-comparison} presents pass counts grouped by difficulty level. 
ORPO ranks third in basic-level tasks with 44/78 (slightly behind Granite-8B-Code-QK at 32/54 and DeepSeek-Coder-33B at 30/54 in terms of completion percentage), achieves the highest pass count on intermediate tasks (41/68), and outperforms others in total completions.  GRPO performs worse than ORPO but still surpasses many baseline models. Neither model succeeds on the five advanced tasks, consistent with all other baselines.

These results offer complementary insights: GRPO appears to be more effective for simpler structurally consistent circuits, benefiting from group-level ranking rewards, while ORPO demonstrates stronger reasoning and robustness on moderately complex tasks due to its fine-grained preference alignment objective.

Both models perform on par with or exceed these reference rates, further validating their practical utility.
Overall, the results highlight the strength of preference-driven optimization in quantum programming and emphasize the importance of evaluating across difficulty levels to capture nuanced model capabilities.
\subsubsection{Training Dynamics}

\begin{figure}[tb]
\centering
\includegraphics[width=0.45\textwidth]{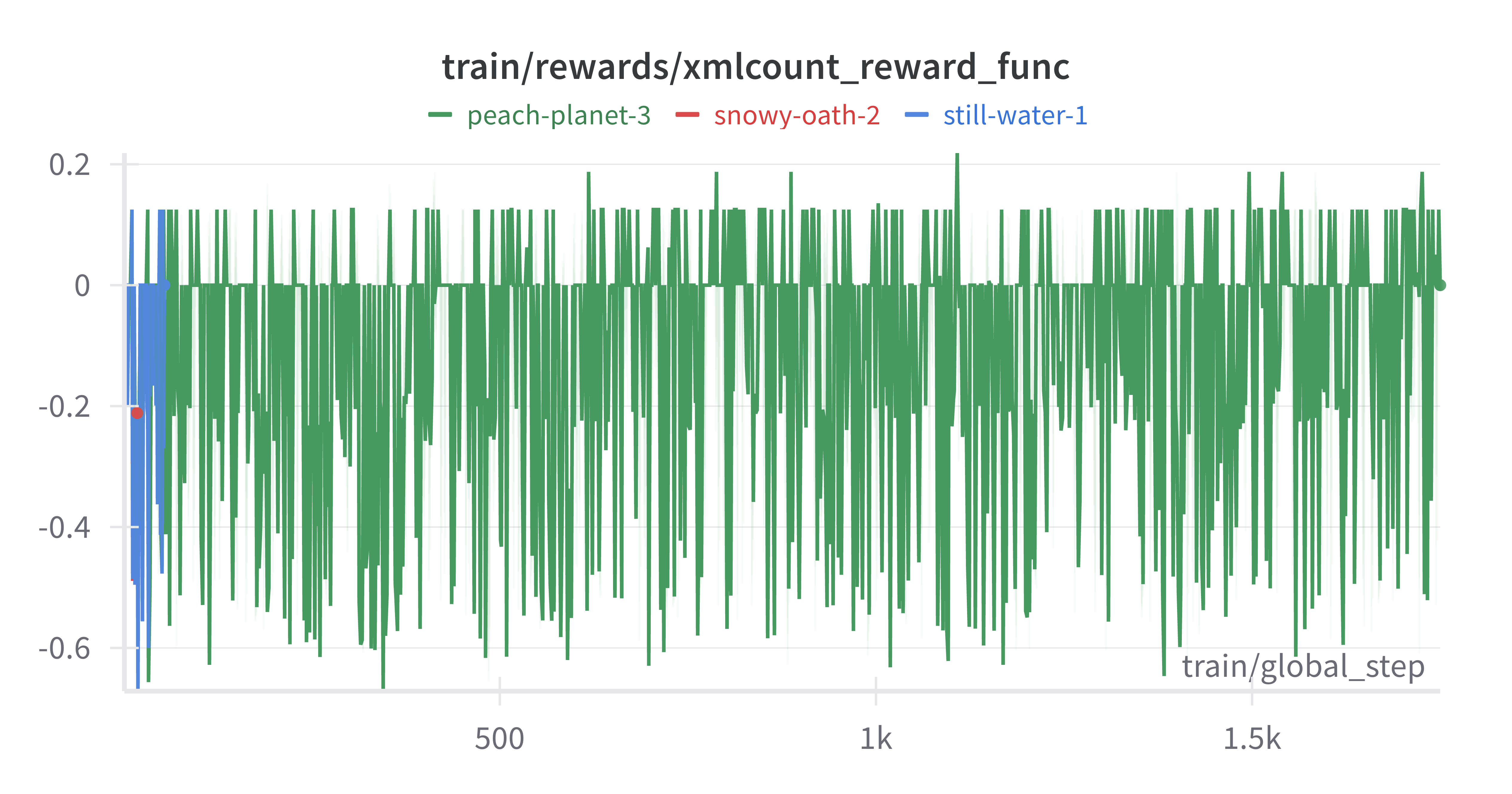}
\includegraphics[width=0.45\textwidth]{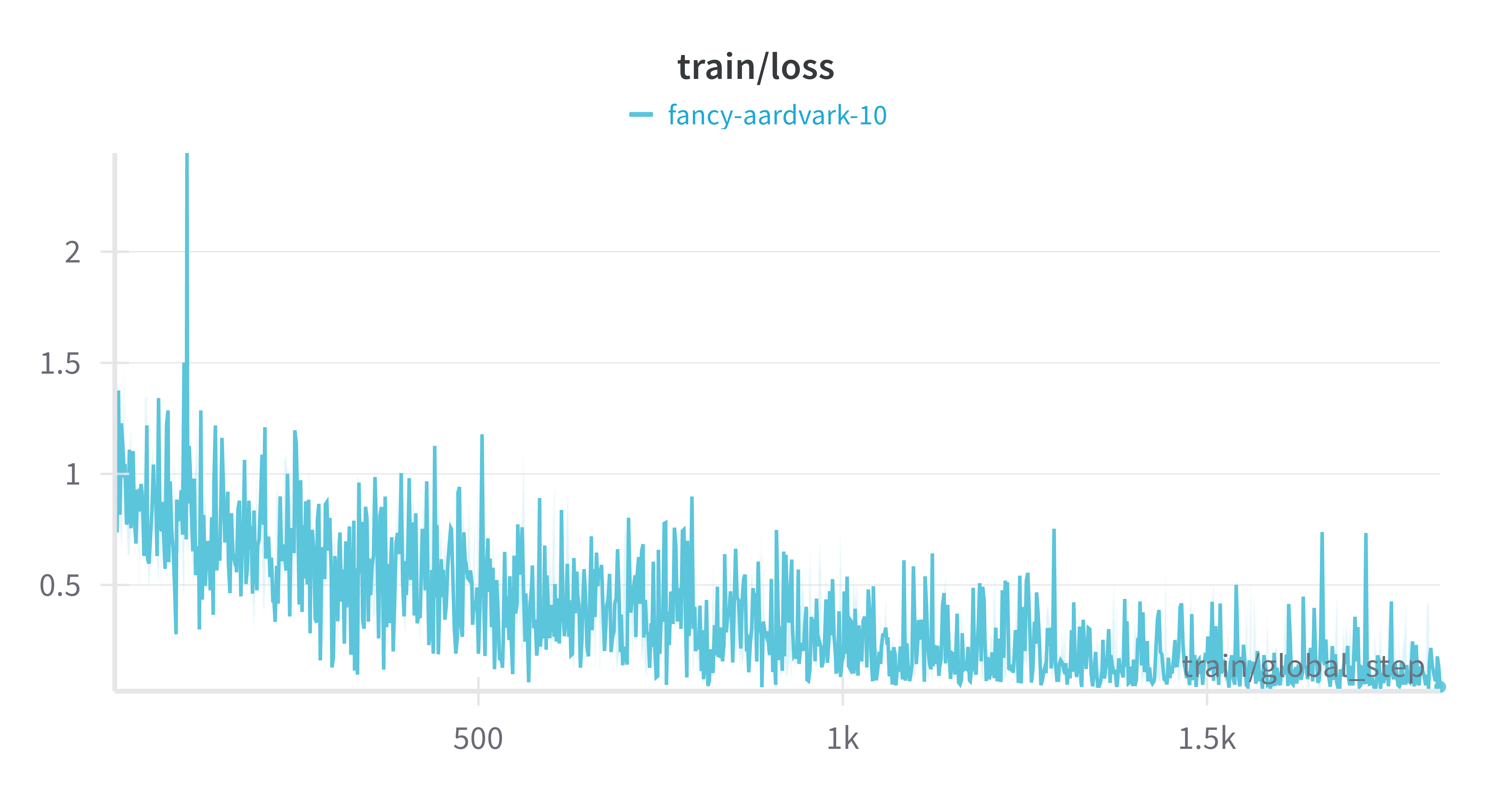}
\caption{Training dynamics of QSpark fine-tuning. 
Left: GRPO reward trajectory with normalized advantage values ($y$-axis) across training steps ($x$-axis). 
Right: ORPO pairwise ranking loss ($y$-axis) across training steps ($x$-axis). 
Legends and axis labels are explicitly provided to highlight reward and loss behavior.}
\label{fig:training-curves}
\end{figure}

Figure~\ref{fig:training-curves} illustrates the training dynamics of our preference-optimized models. The plot on the left shows the reward trajectory for GRPO, while the plot on the right presents the loss curve for ORPO.

In the GRPO setup (left), the model is optimized using group-based reinforcement signals derived from task-specific XML output properties. The observed rewards display high variance throughout training, a result of simulation-based reward assignment and the stochastic nature of quantum program outputs. Despite fluctuations, the trend demonstrates that the model is able to consistently explore and exploit high-reward completions. Importantly, this noisy, yet bounded, reward signal is characteristic of preference-driven reinforcement learning in sparse-reward domains.

The ORPO training curve (right) exhibits more stable and gradual convergence. The pairwise ranking loss steadily decreases as the model learns to align its output with preference-labeled completions. The initial sharp drop is followed by continued fine-tuning and refinement, reflecting effective optimization using contrastive supervision. Huang et al.~\cite{Huang2023Generalization} provide theoretical insights into pairwise learning for ranking, supporting our observation of a smooth convergence process under such loss formulations. This steady progression contrasts with the volatility of GRPO and highlights the complementary strengths of the two methods. GRPO encourages exploration and robustness through output diversity, while ORPO guides the model toward aligning with desired behavior patterns.

Together, these training signals validate the design of our preference learning pipeline. GRPO encourages structural diversity and correctness in simpler tasks, while ORPO promotes nuanced alignment and interpretability in more complex scenarios.

\subsection{Discussion}

Our updated evaluation highlights the effectiveness of reinforcement learning with preferences, through GRPO and ORPO, for quantum code generation. Both models substantially outperform general-purpose LLMs on the Qiskit HumanEval (QHE) benchmark under greedy decoding. ORPO achieves the highest QHE pass@1 score at 56.29\%, surpassing even the domain-specific Granite-8B-Code-QK baseline (46.53\%), while GRPO also delivers a strong 49.00\%. These improvements were achieved without explicit supervised instruction tuning or access to the original QHE fine-tuning scripts, emphasizing the strength of our preference optimization pipeline.
At the task difficulty level, ORPO consistently outperforms GRPO.

Interestingly, even the benchmark reference implementation, executed via the latest version of Qiskit, only achieves 69/78 on Basic, 63/68 on Intermediate, and 2/5 on Advanced tasks. This highlights potential fragility and version sensitivity in quantum execution environments. It also raises a critical point: our models were tested under realistic run-time conditions and still matched or exceeded these reference pass counts, reinforcing their practical applicability.
Compared to previous work by Vishwakarma et al.~\cite{vishwakarma2024qiskit}, our results are competitive despite several evaluation challenges. Although their paper reports 101 tasks, the public release contained 151 files. Furthermore, their evaluation script was not released. To address this, we wrote our benchmark evaluation script and validated completions using the unit tests provided, ensuring consistent and reproducible results.

Finally, the persistent failure across all models, including ours and the benchmark baseline, on the five advanced tasks (0/5) underscores the difficulty of complex quantum reasoning. These results suggest that success in advanced quantum programming likely requires novel strategies, potentially involving curriculum learning, richer supervision signals, or domain-specific memory mechanisms.

Our findings validate reinforcement learning with preferences as a promising direction for quantum LLMs. They also highlight the urgent need for standardized, version-controlled benchmarks and tooling in quantum code generation research.

\subsection{Challenges and Ethical Considerations}

Although the integration of AI into quantum computing offers significant promise, it also introduces important challenges, both technical and ethical, that must be carefully considered ~\cite{IEEEQuantum}.

Lim et al.~\cite{lim2023generative} and Hernandez and Patel \cite{hernandez2025quantumaied} highlight the dual role of generative AI in education, portraying it as both a transformative enabler and a potential threat to traditional learning paradigms. This perspective aligns with our work, where reinforcement learning is used not to replace human quantum programmers but to augment their workflows. Our use of preference optimization explicitly reflects this balance: the goal is to guide AI-generated code toward human-aligned styles, best practices, and interpretable solutions, rather than generate opaque or overly optimized outputs that lack usability.

Ahmadi~\cite{ahmadi2023quantum} explores the convergence of quantum computing and artificial intelligence, emphasizing the revolutionary potential of this union in fields such as cryptography and optimization. However, he also underscores several concerns, ranging from algorithmic reliability to ethical deployment, which are directly relevant to our work. In particular, issues like execution fidelity, qubit resource constraints, and reproducibility are amplified in the quantum domain, where small errors in AI-generated code can lead to significant deviations in output. Our manual validation of test cases, necessitated by the lack of standardized evaluation tools, further reflects the importance of transparency and accountability in quantum AI development.

Looking ahead, we advocate for the development of community-driven benchmarks, shared evaluation pipelines, and stronger documentation practices. These are critical steps not only for reproducibility but also for building trustworthy AI systems that can be safely and ethically deployed in quantum research and education.

\section{Conclusion and Future Work}

In this work, we present a Qiskit-based quantum code assistant built on the Qwen2.5-Coder-32B model, fine-tuned using reinforcement learning with preferences. By introducing Group Relative Policy Optimization (GRPO) and Odds-Ratio Preference Optimization (ORPO), we explore how domain-aligned feedback can improve quantum code generation beyond conventional supervised fine-tuning. Our models demonstrate competitive performance on the Qiskit HumanEval benchmark, particularly excelling at Basic and Intermediate tasks, where they outperform several general-purpose LLMs. These results underscore the promise of preference-based optimization for aligning large language models with quantum programming best practices.

However, this work also presents important challenges. We encountered inconsistencies in benchmark releases, missing evaluation scripts, and had to manually run and validate test cases, which affected reproducibility. Furthermore, none of the evaluated models, including ours, succeeded in the Advanced-level tasks, pointing to the need for better instruction tuning, longer-horizon reasoning, and deeper integration with quantum hardware constraints.
Although our results demonstrate the potential for preference-based optimization for quantum code generation, there are several limitations to be acknowledged. First, the training data remains relatively small compared to classical code datasets, which may restrict generalization to novel quantum tasks. Second, due to the absence of an official evaluation script and inconsistencies in the published benchmark, we relied on manual validation for scoring, introducing potential subjectivity, and making direct comparisons to other models less precise.

In future work, our aim is to:
\begin{itemize}
    \item Integrate GRPO and ORPO into a unified reward framework \cite{Dai2020Queueing}.
    \item Develop sampling-based decoding strategies that align with human-in-the-loop workflows.
    \item Broaden the data set to encompass a wider range of quantum use cases, including error correction, hybrid quantum-classical algorithms, and hardware-specific optimizations.
\end{itemize}

In addition, we intend to work with a more comprehensive and clearly defined benchmark and develop a robust, automated evaluation pipeline to support consistent testing and comparison across models. We also advocate for the open release of standard evaluation tools to support fair benchmarking and collaborative development in quantum LLM research.

By addressing these open challenges, we hope to push the boundaries of AI-assisted quantum programming, making it more accessible, reliable, and aligned with human intent.
\section{Acknowledgements}
This work is partially sponsored by Natural Science and Engineering Research Council of Canada (grants \# 2020-04760 and RGPIN-2022-03886).
\bibliography{sample-ceur}

\begin{thebibliography}{26}
\expandafter\ifx\csname natexlab\endcsname\relax\def\natexlab#1{#1}\fi
\providecommand{\url}[1]{\texttt{#1}}
\providecommand{\href}[2]{#2}
\providecommand{\path}[1]{#1}
\providecommand{\DOIprefix}{doi:}
\providecommand{\ArXivprefix}{arXiv:}
\providecommand{\URLprefix}{URL: }
\providecommand{\Pubmedprefix}{pmid:}
\providecommand{\doi}[1]{\href{http://dx.doi.org/#1}{\path{#1}}}
\providecommand{\Pubmed}[1]{\href{pmid:#1}{\path{#1}}}
\providecommand{\bibinfo}[2]{#2}
\ifx\xfnm\relax \def\xfnm[#1]{\unskip,\space#1}\fi
\bibitem[{Ajagekar et~al.(2019)Ajagekar, Humble, and You}]{Ajagekar2019Quantum}
\bibinfo{author}{A.~Ajagekar}, \bibinfo{author}{T.~Humble}, \bibinfo{author}{F.~You},
\newblock \bibinfo{title}{Quantum computing based hybrid solution strategies for large-scale discrete–continuous optimization problems},
\newblock \bibinfo{journal}{Computers \& Chemical Engineering} \bibinfo{volume}{132} (\bibinfo{year}{2019}) \bibinfo{pages}{106630}. \DOIprefix\doi{10.1016/j.compchemeng.2019.106630}.
\bibitem[{Javadi-Abhari et~al.(2024)Javadi-Abhari, Treinish, Krsulich, Wood, Lishman, Gacon, Martiel, Nation, Bishop, Cross, Johnson, and Gambetta}]{javadiabhari2024qiskit}
\bibinfo{author}{A.~Javadi-Abhari}, \bibinfo{author}{M.~Treinish}, \bibinfo{author}{K.~Krsulich}, \bibinfo{author}{C.~J. Wood}, \bibinfo{author}{J.~Lishman}, \bibinfo{author}{J.~Gacon}, \bibinfo{author}{S.~Martiel}, \bibinfo{author}{P.~D. Nation}, \bibinfo{author}{L.~S. Bishop}, \bibinfo{author}{A.~W. Cross}, \bibinfo{author}{B.~R. Johnson}, \bibinfo{author}{J.~M. Gambetta},
\newblock \bibinfo{title}{Quantum computing with qiskit},
\newblock \bibinfo{journal}{arXiv preprint}  (\bibinfo{year}{2024}). \URLprefix \url{https://arxiv.org/abs/2405.08810}. \DOIprefix\doi{10.48550/arXiv.2405.08810}. \href{http://arxiv.org/abs/2405.08810}{{\tt arXiv:2405.08810}}.
\bibitem[{Dupuis et~al.(2024)Dupuis, Buratti, Vishwakarma, Forrat, Kremer, Faro, Puri, and Cruz-Benito}]{vishwakarma2024qiskit}
\bibinfo{author}{N.~Dupuis}, \bibinfo{author}{L.~Buratti}, \bibinfo{author}{S.~Vishwakarma}, \bibinfo{author}{A.~V. Forrat}, \bibinfo{author}{D.~Kremer}, \bibinfo{author}{I.~Faro}, \bibinfo{author}{R.~Puri}, \bibinfo{author}{J.~Cruz-Benito},
\newblock \bibinfo{title}{Qiskit code assistant: Training llms for generating quantum computing code},
\newblock in: \bibinfo{booktitle}{Proceedings of the IEEE LLM-Aided Design Workshop (LAD)}, \bibinfo{year}{2024}. \URLprefix \url{https://arxiv.org/abs/2405.19495}. \DOIprefix\doi{10.1109/LAD62341.2024.10691762}.
\bibitem[{Vishwakarma et~al.(2024)Vishwakarma, Harkins, Golecha, Bajpe, Dupuis, Buratti, Kremer, Faro, Puri, and Cruz-Benito}]{Qiskit_code_LLM}
\bibinfo{author}{S.~Vishwakarma}, \bibinfo{author}{F.~Harkins}, \bibinfo{author}{S.~Golecha}, \bibinfo{author}{V.~S. Bajpe}, \bibinfo{author}{N.~Dupuis}, \bibinfo{author}{L.~Buratti}, \bibinfo{author}{D.~Kremer}, \bibinfo{author}{I.~Faro}, \bibinfo{author}{R.~Puri}, \bibinfo{author}{J.~Cruz-Benito},
\newblock \bibinfo{title}{Qiskit humaneval: An evaluation benchmark for quantum code generative models},
\newblock in: \bibinfo{booktitle}{Proceedings of the IEEE International Conference on Quantum Computing and Engineering (QCE)}, \bibinfo{year}{2024}, pp. \bibinfo{pages}{1169--1176}. \DOIprefix\doi{10.1109/QCE57912.2024.00001}.
\bibitem[{Murillo et~al.(2025)Murillo, Garcia-Alonso, Moguel, Barzen, Leymann, Ali, Yue, Arcaini, Pérez-Castillo, García-Rodríguez~de Guzmán, Piattini, Ruiz-Cortés, Brogi, Zhao, Miranskyy, and Wimmer}]{murillo2024challenges}
\bibinfo{author}{J.~M. Murillo}, \bibinfo{author}{J.~Garcia-Alonso}, \bibinfo{author}{E.~Moguel}, \bibinfo{author}{J.~Barzen}, \bibinfo{author}{F.~Leymann}, \bibinfo{author}{S.~Ali}, \bibinfo{author}{T.~Yue}, \bibinfo{author}{P.~Arcaini}, \bibinfo{author}{R.~Pérez-Castillo}, \bibinfo{author}{I.~García-Rodríguez~de Guzmán}, \bibinfo{author}{M.~Piattini}, \bibinfo{author}{A.~Ruiz-Cortés}, \bibinfo{author}{A.~Brogi}, \bibinfo{author}{J.~Zhao}, \bibinfo{author}{A.~Miranskyy}, \bibinfo{author}{M.~Wimmer},
\newblock \bibinfo{title}{Quantum software engineering: Roadmap and challenges ahead},
\newblock \bibinfo{journal}{ACM Transactions on Software Engineering and Methodology} \bibinfo{volume}{34} (\bibinfo{year}{2025}). \DOIprefix\doi{10.1145/3712002}.
\bibitem[{Cruz-Benito et~al.(2018)Cruz-Benito, Faro, Mart{\'\i}n-Fern{\'a}ndez, Ther{\'o}n, and Garc{\'\i}a-Pe{\~n}alvo}]{cruz2018deep}
\bibinfo{author}{J.~Cruz-Benito}, \bibinfo{author}{I.~Faro}, \bibinfo{author}{F.~Mart{\'\i}n-Fern{\'a}ndez}, \bibinfo{author}{R.~Ther{\'o}n}, \bibinfo{author}{F.~J. Garc{\'\i}a-Pe{\~n}alvo},
\newblock \bibinfo{title}{A deep-learning-based proposal to aid users in quantum computing programming},
\newblock in: \bibinfo{booktitle}{International Conference on Learning and Collaboration Technologies}, \bibinfo{organization}{Springer}, \bibinfo{publisher}{Springer International Publishing}, \bibinfo{year}{2018}, pp. \bibinfo{pages}{421--430}. \DOIprefix\doi{10.1007/978-3-319-91152-6_32}.
\bibitem[{Bausch and Leditzky(2018)}]{Bausch2018Quantum}
\bibinfo{author}{J.~Bausch}, \bibinfo{author}{F.~Leditzky},
\newblock \bibinfo{title}{Quantum codes from neural networks},
\newblock \bibinfo{journal}{New Journal of Physics} \bibinfo{volume}{22} (\bibinfo{year}{2018}). \DOIprefix\doi{10.1088/1367-2630/ab6cdd}.
\bibitem[{Silva(2018)}]{silva2018practical}
\bibinfo{author}{V.~Silva}, \bibinfo{title}{Practical quantum computing for developers: programming quantum rigs in the cloud using Python, quantum assembly language and IBM QExperience}, \bibinfo{publisher}{Apress}, \bibinfo{year}{2018}.
\bibitem[{Ammermann et~al.(2024)Ammermann, Mauerer, and Schaefer}]{ammermann2024v}
\bibinfo{author}{J.~Ammermann}, \bibinfo{author}{W.~Mauerer}, \bibinfo{author}{I.~Schaefer}, \bibinfo{title}{Towards view-based development of quantum software}, \bibinfo{year}{2024}. \URLprefix \url{https://arxiv.org/abs/2406.18363}. \href{http://arxiv.org/abs/2406.18363}{{\tt arXiv:2406.18363}}.
\bibitem[{Hevia et~al.(2022)Hevia, Peterssen, and Piattini}]{hevia2022quantumpath}
\bibinfo{author}{J.~L. Hevia}, \bibinfo{author}{G.~Peterssen}, \bibinfo{author}{M.~Piattini},
\newblock \bibinfo{title}{Quantumpath: A quantum software development platform},
\newblock \bibinfo{journal}{Software: Practice and Experience} \bibinfo{volume}{52} (\bibinfo{year}{2022}) \bibinfo{pages}{1517--1530}.
\bibitem[{Liang et~al.(2023)Liang, Cheng, Yang, Ren, Song, Wu, Qian, Li, and Shi}]{liang2023unleashing}
\bibinfo{author}{Z.~Liang}, \bibinfo{author}{J.~Cheng}, \bibinfo{author}{R.~Yang}, \bibinfo{author}{H.~Ren}, \bibinfo{author}{Z.~Song}, \bibinfo{author}{D.~Wu}, \bibinfo{author}{X.~Qian}, \bibinfo{author}{T.~Li}, \bibinfo{author}{Y.~Shi},
\newblock \bibinfo{title}{Unleashing the potential of llms for quantum computing: A study in quantum architecture design},
\newblock \bibinfo{journal}{arXiv preprint arXiv:2307.08191}  (\bibinfo{year}{2023}).
\bibitem[{Aragonés-Soria and Oriol(2024)}]{c4q}
\bibinfo{author}{Y.~Aragonés-Soria}, \bibinfo{author}{M.~Oriol},
\newblock \bibinfo{title}{C4q: A chatbot for quantum},
\newblock in: \bibinfo{booktitle}{Proceedings of the 5th ACM/IEEE International Workshop on Quantum Software Engineering}, Q-SE 2024, \bibinfo{publisher}{Association for Computing Machinery}, \bibinfo{address}{New York, NY, USA}, \bibinfo{year}{2024}, p. \bibinfo{pages}{29–36}. \URLprefix \url{https://doi.org/10.1145/3643667.3648222}. \DOIprefix\doi{10.1145/3643667.3648222}.
\bibitem[{Hong et~al.(2024)Hong, Lee, and Thorne}]{hong2024orpomonolithicpreferenceoptimization}
\bibinfo{author}{J.~Hong}, \bibinfo{author}{N.~Lee}, \bibinfo{author}{J.~Thorne}, \bibinfo{title}{Orpo: Monolithic preference optimization without reference model}, \bibinfo{year}{2024}. \URLprefix \url{https://arxiv.org/abs/2403.07691}. \href{http://arxiv.org/abs/2403.07691}{{\tt arXiv:2403.07691}}.
\bibitem[{Shao et~al.(2024)Shao, Wang, Zhu, Xu, Song, Bi, Zhang, Zhang, Li, Wu, and Guo}]{shao2024deepseekmathpushinglimitsmathematical}
\bibinfo{author}{Z.~Shao}, \bibinfo{author}{P.~Wang}, \bibinfo{author}{Q.~Zhu}, \bibinfo{author}{R.~Xu}, \bibinfo{author}{J.~Song}, \bibinfo{author}{X.~Bi}, \bibinfo{author}{H.~Zhang}, \bibinfo{author}{M.~Zhang}, \bibinfo{author}{Y.~Li}, \bibinfo{author}{Y.~Wu}, \bibinfo{author}{D.~Guo}, \bibinfo{title}{Deepseekmath: Pushing the limits of mathematical reasoning in open language models}, \bibinfo{year}{2024}. \URLprefix \url{https://arxiv.org/abs/2402.03300}. \href{http://arxiv.org/abs/2402.03300}{{\tt arXiv:2402.03300}}.
\bibitem[{et~al.(2021)}]{chen2021evaluatinglargelanguagemodels}
\bibinfo{author}{M.~C. et~al.}, \bibinfo{title}{Evaluating large language models trained on code}, \bibinfo{year}{2021}. \URLprefix \url{https://arxiv.org/abs/2107.03374}. \href{http://arxiv.org/abs/2107.03374}{{\tt arXiv:2107.03374}}.
\bibitem[{Rozière et~al.(2024)Rozière, Gehring, Gloeckle, Sootla, Gat, Tan, Adi, Liu, Sauvestre, Remez, Rapin, Kozhevnikov, Evtimov, Bitton, Bhatt, Ferrer, Grattafiori, Xiong, Défossez, Copet, Azhar, Touvron, Martin, Usunier, Scialom, and Synnaeve}]{rozière2024codellamaopenfoundation}
\bibinfo{author}{B.~Rozière}, \bibinfo{author}{J.~Gehring}, \bibinfo{author}{F.~Gloeckle}, \bibinfo{author}{S.~Sootla}, \bibinfo{author}{I.~Gat}, \bibinfo{author}{X.~E. Tan}, \bibinfo{author}{Y.~Adi}, \bibinfo{author}{J.~Liu}, \bibinfo{author}{R.~Sauvestre}, \bibinfo{author}{T.~Remez}, \bibinfo{author}{J.~Rapin}, \bibinfo{author}{A.~Kozhevnikov}, \bibinfo{author}{I.~Evtimov}, \bibinfo{author}{J.~Bitton}, \bibinfo{author}{M.~Bhatt}, \bibinfo{author}{C.~C. Ferrer}, \bibinfo{author}{A.~Grattafiori}, \bibinfo{author}{W.~Xiong}, \bibinfo{author}{A.~Défossez}, \bibinfo{author}{J.~Copet}, \bibinfo{author}{F.~Azhar}, \bibinfo{author}{H.~Touvron}, \bibinfo{author}{L.~Martin}, \bibinfo{author}{N.~Usunier}, \bibinfo{author}{T.~Scialom}, \bibinfo{author}{G.~Synnaeve}, \bibinfo{title}{Code llama: Open foundation models for code}, \bibinfo{year}{2024}. \URLprefix \url{https://arxiv.org/abs/2308.12950}. \href{http://arxiv.org/abs/2308.12950}{{\tt arXiv:2308.12950}}.
\bibitem[{DeepSeek-AI(2025)}]{deepseekai2025deepseekr1incentivizingreasoningcapability}
\bibinfo{author}{DeepSeek-AI}, \bibinfo{title}{Deepseek-r1: Incentivizing reasoning capability in llms via reinforcement learning}, \bibinfo{year}{2025}. \URLprefix \url{https://arxiv.org/abs/2501.12948}. \href{http://arxiv.org/abs/2501.12948}{{\tt arXiv:2501.12948}}.
\bibitem[{et~al(2023)}]{li2023starcodersourceyou}
\bibinfo{author}{R.~L. et~al}, \bibinfo{title}{Starcoder: may the source be with you!}, \bibinfo{year}{2023}. \URLprefix \url{https://arxiv.org/abs/2305.06161}. \href{http://arxiv.org/abs/2305.06161}{{\tt arXiv:2305.06161}}.
\bibitem[{Team(2024)}]{codegemmateam2024codegemmaopencodemodels}
\bibinfo{author}{C.~Team}, \bibinfo{title}{Codegemma: Open code models based on gemma}, \bibinfo{year}{2024}. \URLprefix \url{https://arxiv.org/abs/2406.11409}. \href{http://arxiv.org/abs/2406.11409}{{\tt arXiv:2406.11409}}.
\bibitem[{Mishra et~al.(2024)Mishra, Stallone, Zhang, Shen, Prasad, Soria, Merler, Selvam, Surendran, Singh, Sethi, Dang, Li, Wu, Zawad, Coleman, White, Lewis, Pavuluri, Koyfman, Lublinsky, de~Bayser, Abdelaziz, Basu, Agarwal, Zhou, Johnson, Goyal, Patel, Shah, Zerfos, Ludwig, Munawar, Crouse, Kapanipathi, Salaria, Calio, Wen, Seelam, Belgodere, Fonseca, Singhee, Desai, Cox, Puri, and Panda}]{mishra2024granitecodemodelsfamily}
\bibinfo{author}{M.~Mishra}, \bibinfo{author}{M.~Stallone}, \bibinfo{author}{G.~Zhang}, \bibinfo{author}{Y.~Shen}, \bibinfo{author}{A.~Prasad}, \bibinfo{author}{A.~M. Soria}, \bibinfo{author}{M.~Merler}, \bibinfo{author}{P.~Selvam}, \bibinfo{author}{S.~Surendran}, \bibinfo{author}{S.~Singh}, \bibinfo{author}{M.~Sethi}, \bibinfo{author}{X.-H. Dang}, \bibinfo{author}{P.~Li}, \bibinfo{author}{K.-L. Wu}, \bibinfo{author}{S.~Zawad}, \bibinfo{author}{A.~Coleman}, \bibinfo{author}{M.~White}, \bibinfo{author}{M.~Lewis}, \bibinfo{author}{R.~Pavuluri}, \bibinfo{author}{Y.~Koyfman}, \bibinfo{author}{B.~Lublinsky}, \bibinfo{author}{M.~de~Bayser}, \bibinfo{author}{I.~Abdelaziz}, \bibinfo{author}{K.~Basu}, \bibinfo{author}{M.~Agarwal}, \bibinfo{author}{Y.~Zhou}, \bibinfo{author}{C.~Johnson}, \bibinfo{author}{A.~Goyal}, \bibinfo{author}{H.~Patel}, \bibinfo{author}{Y.~Shah}, \bibinfo{author}{P.~Zerfos}, \bibinfo{author}{H.~Ludwig}, \bibinfo{author}{A.~Munawar}, \bibinfo{author}{M.~Crouse}, \bibinfo{author}{P.~Kapanipathi},
  \bibinfo{author}{S.~Salaria}, \bibinfo{author}{B.~Calio}, \bibinfo{author}{S.~Wen}, \bibinfo{author}{S.~Seelam}, \bibinfo{author}{B.~Belgodere}, \bibinfo{author}{C.~Fonseca}, \bibinfo{author}{A.~Singhee}, \bibinfo{author}{N.~Desai}, \bibinfo{author}{D.~D. Cox}, \bibinfo{author}{R.~Puri}, \bibinfo{author}{R.~Panda}, \bibinfo{title}{Granite code models: A family of open foundation models for code intelligence}, \bibinfo{year}{2024}. \URLprefix \url{https://arxiv.org/abs/2405.04324}. \href{http://arxiv.org/abs/2405.04324}{{\tt arXiv:2405.04324}}.
\bibitem[{Huang et~al.(2023)Huang, Zhou, Feng, and Zhou}]{Huang2023Generalization}
\bibinfo{author}{S.~Huang}, \bibinfo{author}{J.~Zhou}, \bibinfo{author}{H.~Feng}, \bibinfo{author}{D.-X. Zhou},
\newblock \bibinfo{title}{Generalization analysis of pairwise learning for ranking with deep neural networks},
\newblock \bibinfo{journal}{Neural Computation} \bibinfo{volume}{35} (\bibinfo{year}{2023}) \bibinfo{pages}{1135--1158}. \DOIprefix\doi{10.1162/neco_a_01585}.
\bibitem[{Rawat and Bajracharya(2024)}]{IEEEQuantum}
\bibinfo{author}{D.~B. Rawat}, \bibinfo{author}{C.~Bajracharya},
\newblock \bibinfo{title}{The intersection of quantum computing, ai, and cybersecurity: Challenges and opportunities},
\newblock in: \bibinfo{booktitle}{2024 IEEE 6th International Conference on Trust, Privacy and Security in Intelligent Systems, and Applications (TPS-ISA)}, \bibinfo{year}{2024}, pp. \bibinfo{pages}{176--181}. \DOIprefix\doi{10.1109/TPS-ISA62245.2024.00029}.
\bibitem[{Lim et~al.(2023)Lim, Gunasekara, Pallant, Pallant, and Pechenkina}]{lim2023generative}
\bibinfo{author}{W.~M. Lim}, \bibinfo{author}{A.~Gunasekara}, \bibinfo{author}{J.~L. Pallant}, \bibinfo{author}{J.~I. Pallant}, \bibinfo{author}{E.~Pechenkina},
\newblock \bibinfo{title}{Generative ai and the future of education},
\newblock \bibinfo{journal}{The International Journal of Management Education}  (\bibinfo{year}{2023}). \URLprefix \url{https://www.sciencedirect.com/science/article/pii/S1472811723000289}. \DOIprefix\doi{10.1016/j.ijme.2023.100790}.
\bibitem[{Hernandez and Patel(2025)}]{hernandez2025quantumaied}
\bibinfo{author}{K.~Hernandez}, \bibinfo{author}{T.~Patel},
\newblock \bibinfo{title}{Enhancing early quantum computing education with quantumaied: Bridging the educational gap},
\newblock in: \bibinfo{booktitle}{Proceedings of the ACM SIGCSE}, \bibinfo{organization}{ACM}, \bibinfo{year}{2025}, p. \bibinfo{pages}{1755}. \URLprefix \url{https://dl.acm.org/doi/10.1145/3641555.3705028}.
\bibitem[{Ahmadi(2023)}]{ahmadi2023quantum}
\bibinfo{author}{A.~Ahmadi},
\newblock \bibinfo{title}{Quantum computing and artificial intelligence: The synergy of two revolutionary technologies},
\newblock \bibinfo{journal}{American Journal of Engineering and Science (AJES)}  (\bibinfo{year}{2023}). \URLprefix \url{https://ajesjournal.org/index.php/ajes/article/view/4118}. \DOIprefix\doi{10.51983/ajes-2023.12.2.4118}.
\bibitem[{Dai and Gluzman(2020)}]{Dai2020Queueing}
\bibinfo{author}{J.~Dai}, \bibinfo{author}{M.~O. Gluzman},
\newblock \bibinfo{title}{Queueing network controls via deep reinforcement learning},
\newblock \bibinfo{journal}{ArXiv} \bibinfo{volume}{abs/2008.01644} (\bibinfo{year}{2020}). \DOIprefix\doi{10.1287/stsy.2021.0081}.

\end{thebibliography}

\end{document}